\documentclass[11pt,twoside]{article}
\usepackage{asp2010}
\usepackage{natbib}
\newcommand{\kms}{\mbox{km~s$^{-1}$}}

\resetcounters

\begin{document}

\title{What collisional debris can tell us about galaxies}
\author{Pierre-Alain Duc,$^1$} 
\affil{$^1$ AIM Paris-Saclay, CNRS/INSU, CEA/Irfu, Universit\'e Paris Diderot, CEA-Saclay, Orme des merisiers, 91191 Gif sur Yvette cedex, France}

\begin{abstract}
I review what tidal tails in particular, collisional debris in general, might tell us about galaxies (their structure, current content and past mass assembly) about mergers in the nearby and distant Universe (major vs minor, wet vs dry, number evolution) and finally about the laws of gravity. 

\end{abstract}

\section{Introduction,  words of caution and acknowledgments}
{\it Tidal tails} are one of the most  visible upshots of on-going or past galaxy collisions. 
They  consist of material that has been expelled from {\it parent} galaxies under the effect of gravity and {\it tides}. Tidal tails are formed early on during the collision, soon after the first passage. They subsist long after the merger, though evaporate with time or fall back onto their progenitors. 
Tidal tails are part of the  debris generated by collisions together with {\it bridges, plumes, shells and rings}.

 The reader is referred to \citet{Duc11c} for an extensive overview on tides in galaxies. It presents a historical perspective, the theoretical, numerical and observational approaches so far used to  study them, and gives a detailed bibliography.
  The ambition of this Review  is much more restricted. Emphasizing on basic, over-simplified, facts about collisional debris,  the Paper  may be entitled  with no desire to offend the reader, ``Tidal tails for Dummies''.  Thus, the bibliography has on purpose  been ignored;  the way some results are presently is certainly biased. One of the motivation put forward by the organizers for  convincing participants to attend the meeting was that Proceeding papers do not have referees. These words  were taken seriously here. 
The original accent of the Presenter, noted in the concluding remarks of the conference, may still be present in his written version. The mercy of the reader is asked for that. 
 Finally the Writer takes the opportunity of this introduction to thank the Local Organizing Committee for the quasi-perfect organization of the conference.
  
  The paper is organized as follows: after having presented what tidal tails may tell us about the nature of mergers, it discusses what tidal tails may tell us about the structure and  content of galaxies. It then presents what tidal tails may tell us about the laws of gravity, before addressing what tidal tails may tell us about the mass assembly of galaxies. Finally it concludes in questioning what really tidal tails may tell us  about the Universe.

\section{What tidal tails can tell us about mergers}
\subsection{Merger type: major vs minor}
\label{sect:maj-min}
The number and morphology of the tidal  tails depend on the mass ratio between the parent galaxies.

Tidal tails that result from {\it major} mergers between spiral galaxies 
 are long --  typically 20-100 kpc --,  curved and binary. A main tail  and a counter-tail  are formed from each colliding galaxy. 
At the post-merger phase, the two counter-tails have already disappeared;  remain the two main tails.  
A caricatural  example of such a merger may be seen in Figure~\ref{fig:major}.

\articlefigure[width=\textwidth]{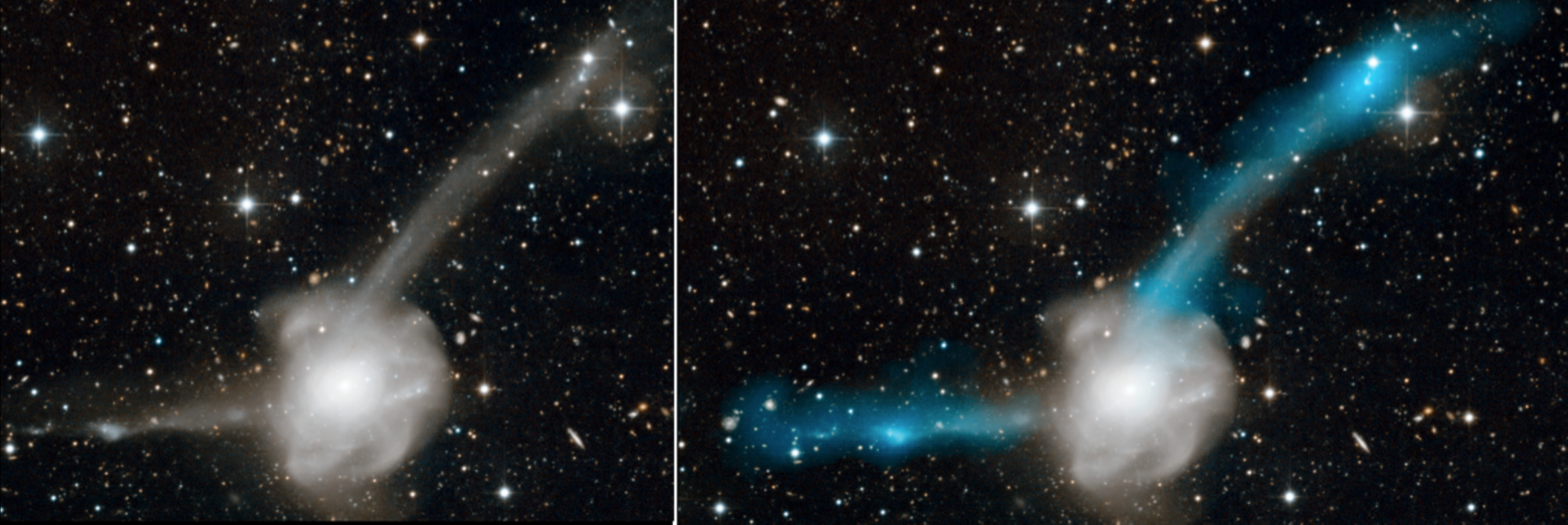}{fig:major}{True-color ESO optical image of the prototypical merger NGC~7252 with, to the right,  a  NRAO/VLA HI map superimposed.}

In reality, tidal tails are 3D structures. Projection effects might result in complex morphologies. Tidal effects during multiple passages further contribute to blur the story. Tidal material falling back generate  {\it loops}.

In a {\it minor} merger, only the smaller companion is significantly tidally disrupted. Two long and narrow tails form, making a small angle. Joined together, the major/minor tail might be mistaken with a single major tail. The presence of a prominent stellar knot somewhere along the tail - the remnant of the disrupted satellite - reveals the real nature of the system. Tidal material wraps around the major galaxy, while the satellite continues to be stripped, until it is totally destroyed. 
An example of tidal tails generated by a minor collision is presented in  Figure~\ref{fig:minor} (left).

\articlefigure[width=10cm]{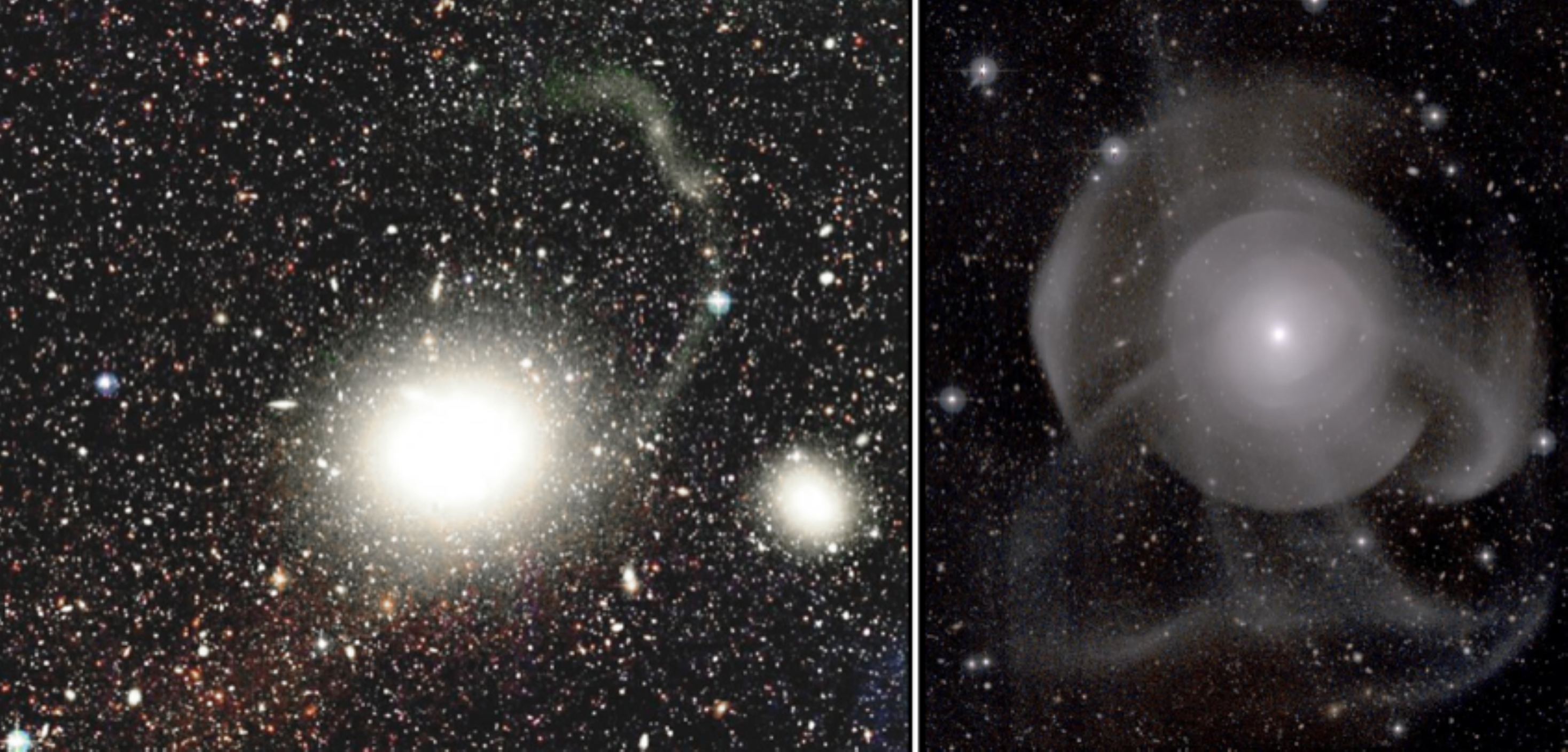}{fig:minor}{Remnants of minor mergers on CFHT deep optical images. {\it Left}: tidal tails from a disrupted satellite in the Next Generation Virgo Cluster Survey\protect\footnotemark[1]. {\it Right}: a network of shells in an early-type galaxy from the ATLAS$^{3D}$ sample\protect\footnotemark[2].}

\subsection{Merger orbits: prograde vs retrograde, small and large impact parameters}
The way the two galaxies encounter has an impact on the shape of the tidal tails.
A {\it prograde} orbit generates long, narrow,  tidal tails. Conversely, a more diffuse, shorter, {\it plume} like structure,  suggests a  {\it retrograde} orbit.

If the {\it impact} parameter diminishes, a {\it ring} is   formed  (See Figure~\ref{fig:ring}).  It  expands with time, together with the density waves at its origin. In the case of minor mergers, {\it shells} are formed instead. These sharp-edged arcs might be numerous and more or less  concentric (see Figure~\ref{fig:minor}, right). 
The number of shells tell us about the  initial  energy of the collision. 

\articlefigure[width=10cm]{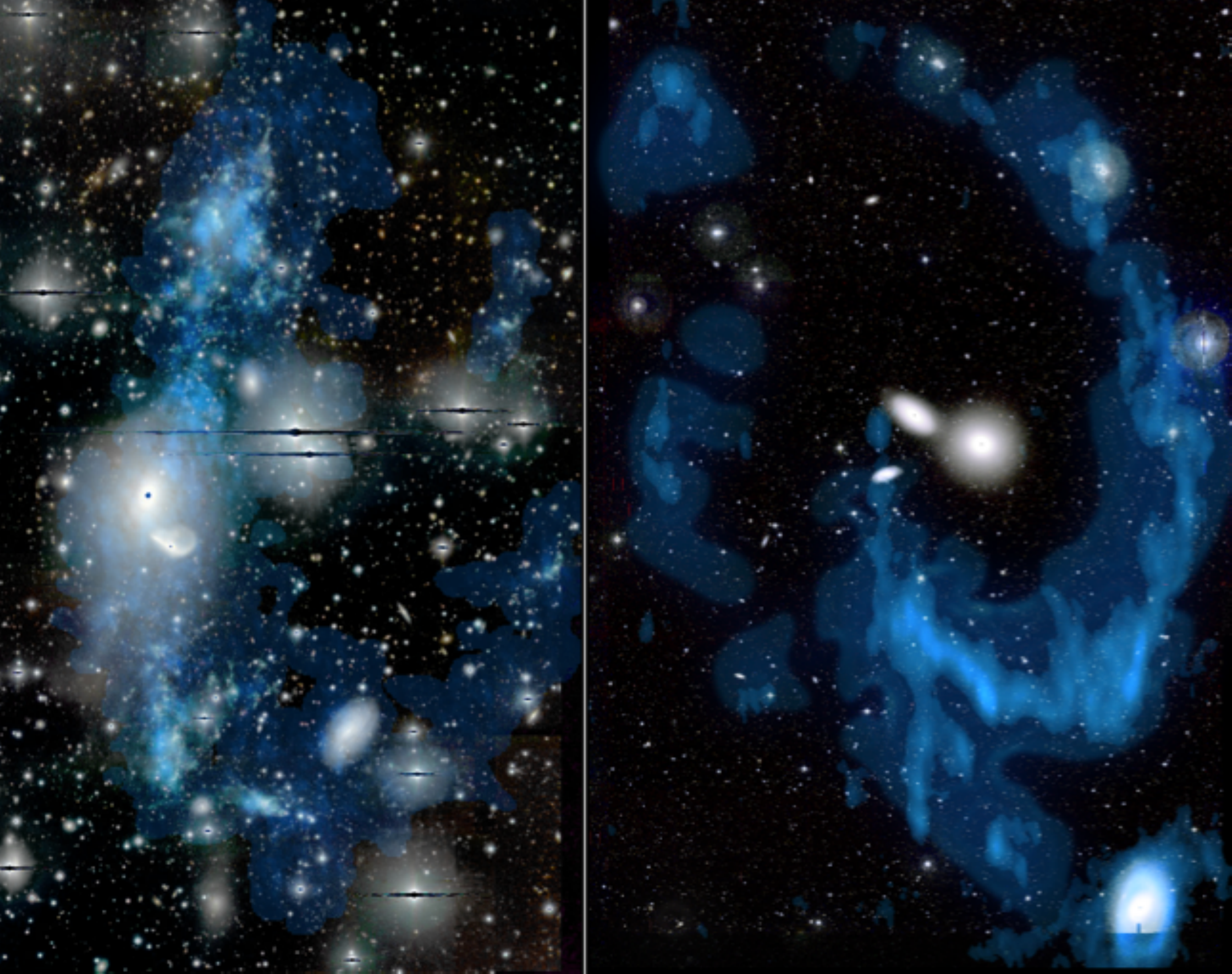}{fig:ring}{Two giant gaseous rings formed after nearly head-on collisions with massive companions. {\it Left}: NGC~5291; {\it Right}: The Leo Ring. VLA/WSRT HI maps are superimposed on ESO/CFHT true color images.}

\footnotetext[1]{https://www.astrosci.ca/NGVS/The\_Next\_Generation\_Virgo\_Cluster\_Survey/}
\footnotetext[2]{http://www-astro.physics.ox.ac.uk/atlas3d/}

\subsection{Velocity speed: slow merger vs fast fly-by}

The velocity with which the two galaxies encounter, and thus the duration of the collision,  have an impact on the development of  tidal tails.
Above 1000~\kms, tidal perturbations are minimum. Galaxies are slightly perturbed, and only repeated high-speed {\it fly-bys}  might have an impact on their morphology, a process known in clusters  as {\it harassment}. 

Effects may be more dramatic  if the galaxy flying-by  is very massive. Part of the least gravitationally bound component of the target -- the HI gas --   is  expelled,
while the bullet is unaffected  and continues its journey. 

\subsection{Merger age}
One of the goals of idealized numerical simulations of mergers is to reproduce the current shape (and kinematics) of collisional debris. If the observed and simulated structures match with a given  set of parameters, this solution is kept as a likely one, and the age of the merger is determined.
The morphology of tidal tails  and the support of a numerical model thus provide a time clock for dating events associated with the merger, such as starbursts.

\section{What tidal tails can tell us about the structure of galaxies}
The ability of a collision to produce collisional debris does also depend on  the structural parameters of the colliding galaxies.

\subsection{Morphology: late-type vs early-type progenitors}
\label{sect:morph}
In the case of a major merger, the presence of a tidal tail requires that at least one of the parent galaxy had a disk. Indeed, it is extremely difficult to expel material from a hot system, such as a spheroid. 
In other words, a collision between two ellipticals will not produce any visible tidal structure, as illustrated on Figure~4. Only  weak asymmetries in the stellar halo might be induced.

\begin{figure}[h]
   \begin{center}
    \begin{minipage}[t]{0.5\linewidth}
  \raisebox{-2.5cm}{  \includegraphics[width=7cm, angle=0]{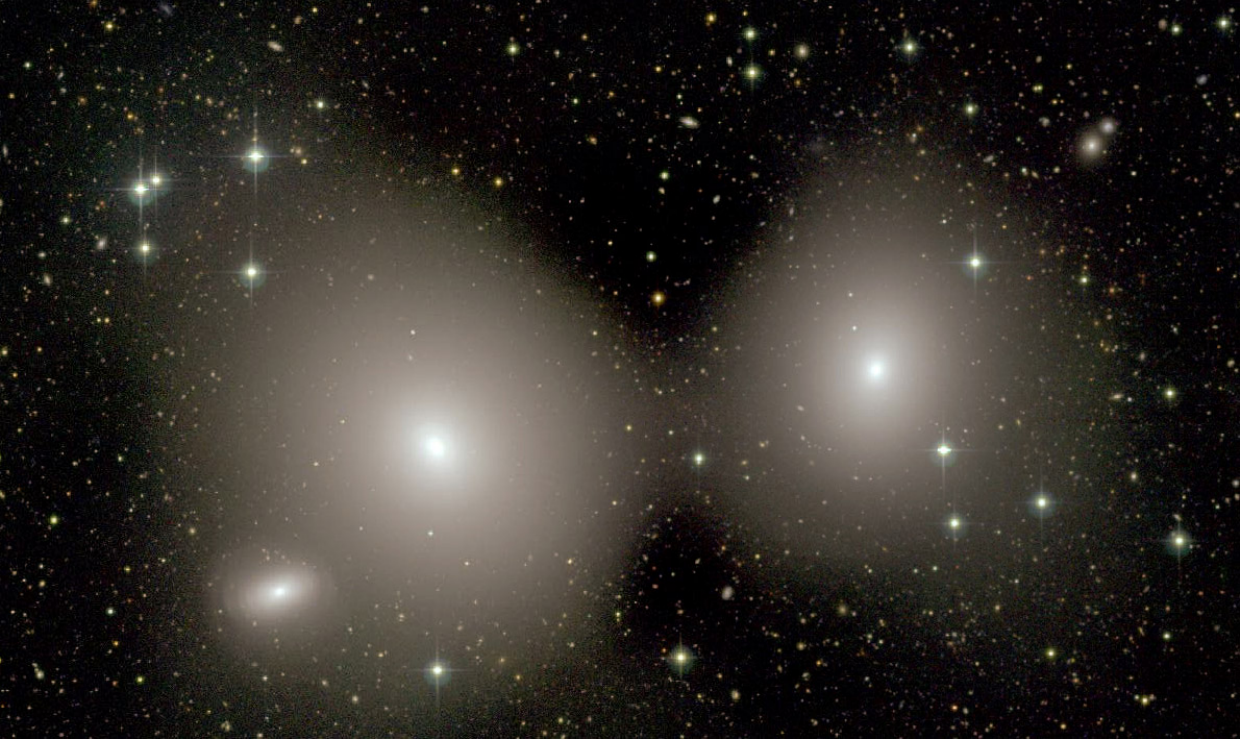}}
  \end{minipage} \hfill
   \begin{minipage}[t]{0.5\linewidth}
 \caption{Example of a {\it dry} merger of two  ATLAS$^{3D}$ ellipticals. Note the absence of tidal tails on these deep CFHT true color images.}
\end{minipage}
\end{center}
\label{fig:early}
\end{figure}

In the case of minor mergers though, the tidal shock  induced by the massive host is such that even rather hot systems like early-type satellites or globular clusters are disrupted and stellar streams are produced. 

\subsection{Dark matter: truncated vs extended halo}
\label{Sect:halo}
Because tidal tails of major mergers have sizes reaching that of the dark-matter (DM) halo of their parent, they may in principle be used as a probe of the total  DM content of galaxies. Intuition tells that an extended, massive, DM halo would hamper the formation of tidal tails. Numerical simulations state that the length of the tail is not directly linked to the size of the halo. In fact, the radial distribution and concentration of the DM particles  matter more. 
The ability of a tail to develop in situ massive sub-structures depends itself  on the extent of the DM halo: massive tidal condensations such as the observed {\it Tidal Dwarf Galaxies} are not formed if the halo is truncated. 
Finally, modeling the morphology and orbits of stellar streams of disrupted satellites  allows to investigate the shape of the DM halo of the host galaxy, in particular its flatness and tri-axiality.

\section{What tidal tails can tell us about the content of galaxies}

Collisional debris is like a  trash bin: determine its content and you learn about its owner possession and way of life.
Interacting galaxies are just like giant colliders: detect the particles/structures created by the collision and you  learn about the structure of matter/galaxies.

\subsection{HI  gas: wet versus dry mergers}
The literature is prolific in discussions whether the mass assembly of galaxies is done through {\it wet},  gas--rich mergers, i.e. one of the parent galaxy contained some gas,  or {\it dry}, gas--poor, mergers,  i.e. both parent galaxies were gas poor, i.e. already {\it red and dead}. 
If one of the parent galaxies hosted an extended HI disk, gas should also be present in the tidal tails. Radio observations show this is the case. Atomic hydrogen is in fact the most massive constituent of tidal tails  generated by major mergers (See  Figure~\ref{fig:major}). 
However when the merger gets old, gas  is dispersed and  no longer visible. At intermediate and high redshift, it is  even not observable with current facilities. Remember however that if  a stellar tail is visible, its parent had a stellar disk, i.e.  it was  gas-rich. Thus, the sole detection of tidal tails implies that the merger was wet, even if  the remnant already looks red and dead.

\subsection{Molecular gas: visible vs dark gas}
\label{Sect:dark}

Not only atomic gas is found in collisional debris. Surprisingly large quantities of molecular gas, traced by  CO, and associated with dust (that emits in the mid-IR/far-IR), were detected there. Is however molecular gas formed locally, within the densest HI clouds in the tidal tails or was it expelled from the parent galaxies, together with the stars and HI? Probably both: if the bulk of the CO emission is generally measured towards the main HI condensation, a more diffuse, 'hotter' , i.e. with wider lines and thus larger velocity dispersion, is found all along the tails of the few systems that could be fully mapped in CO. 

Within tidal tails, some HI clouds are so massive and dense that they become gravitationally bound and start rotating, offering an avenue to determine their total mass. Their luminous - visible mass - being known from optical, HI and CO observations, the difference is due to the dark component. The few measures so far made on presumably bound gas condensations tend to indicate that a dark component is present in the debris.... and thus in the disk of their progenitors. Since only very small quantities of non-baryonic dark matter is present in disks, the dark component of collisional  debris should be dark baryons. The most likely hypothesis is invisible molecular gas, i.e. not traced by CO, and recently found in nearby spirals, including our own Milky Way, by observations of the dust  component.
Collisional debris of wet, major, mergers offer an independent   laboratory to check the darkest constituents of galaxies.

\subsection{Stars: young versus old ones}

Collisional stellar debris are colorful structures: red from their old stars, and blue from the stars born in situ, when the atomic/molecular gas clouds mentioned above collapse. The ratio between the young and old components, derived from analyzing the Spectral Energy Distribution, likely depends on the gas content of the tails, and thus of the parent galaxies.

Collisional debris are  a special environment: outside a prominent stellar disk, but with the interstellar medium (gas and  dust) of stellar disks. Studying how star-formation proceeds there will tell us how star-formation proceeds in general. No big surprise has yet been found, with star-formation occurring in tidal tails  just like in spiral disks. Therefore tidal tails  are not guilty for the deviant behavior of the star-formation laws claimed in nearby and distant interacting systems.

\section{What tidal tails can tell us about physics}

Tidal debris may even probe the hidden secrets of physics, and as astrophysicists are concerned, the most intriguing one: gravity. 

The dynamical status of some bound clumps has been addressed in Sect.~\ref{Sect:dark} where it was said that their kinematics implied the (unexpected) presence of a dark component. As a matter of fact, or belief, rotation curves of such structures (that qualify as {\it Tidal Dwarf Galaxies}) may be reproduced without this extra component, provided that gravity is modified at small acceleration parameters. MOND can fit all data, even those given by collisional debris, ... within the error-bars. Unfortunately they are still too large to claim anything securely.
Would the resolution of the kinematical data increase, collisional debris have the potential  to  become a MOND-supporter ... or killer.

\section{What tidal tails can tell us about the formation of galaxies}
We have now all the arguments to discuss what tidal tails might tell us on how galaxies are born, grow and age.

Let us start with the birth and with what might be an anecdote  or not:  the evidence for the formation of second generation of galaxies  in collisional debris, the already mentioned baby objects of the Author.   How many  {\it Tidal Dwarf Galaxies} are born in mergers and survive is very debated (but not in this review).

\subsection{Mass assembly of galaxies: from the past to the present: observational cosmology}

\articlefigure[width=\textwidth]{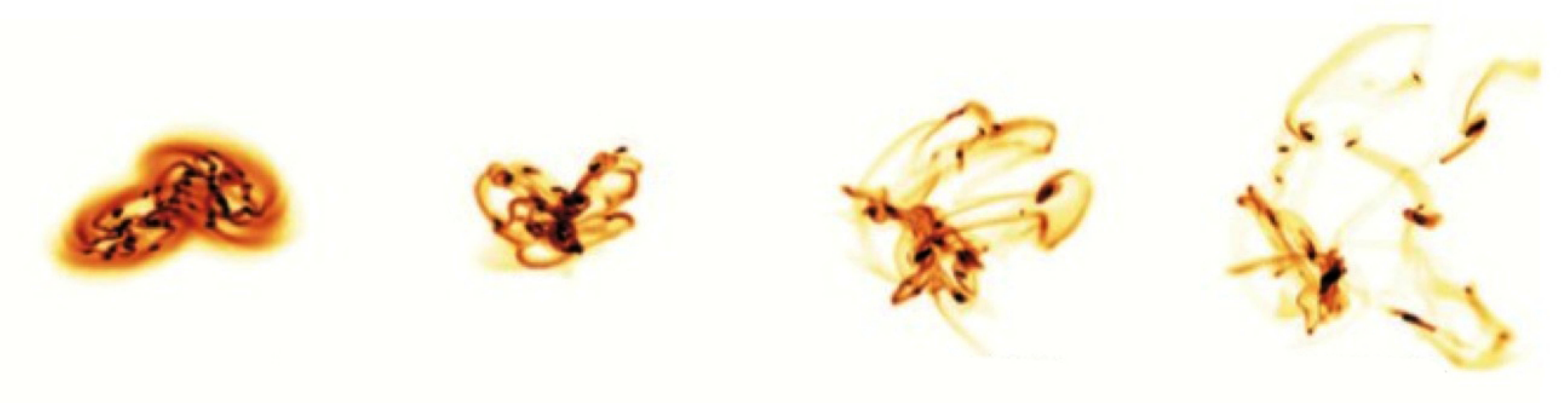}{fig:distmerg}{Numerical simulations of a merger between two distant  clumpy galaxies. Courtesy F. Bournaud}

The presence of a tidal tail appears as an unambiguous proof of a past merger. So in principle, tidal perturbations may be used to estimate the evolution of the merger rate as a function of redshift,  a key parameter in the $\Lambda$-CDM hierarchical scenario: galaxies grow through multiple minor vs major, wet vs dry mergers.
In the past, the Universe was denser; collisions were more numerous. Thus the merger rate should increase with redshift, such as the number of pairs -- the other way to infer the evolution of the merger rate --  and distant galaxies appear much more disturbed than now. 
Having said that, the reader is invited to look at Figure~\ref{fig:distmerg}, presenting an example of a simulated high-redshift collision. The progenitors are typical for $z=2$: very clumpy, due to their barely stable gaseous disk. Even before the collision, the galaxies look disturbed. One may associate  each clump with a galactic nucleus and  conclude that the system  is a multiple merger.  However, a kinematical analysis indicates that these objects are in fact rotating disks, and the  clumps giant star-forming regions, with masses typical of dwarf galaxies. Now, when two such clumpy galaxies interact, making a real merger, it is remarkable that contrary to what is observed in nearby mergers, no big  tidal tails develop\footnote{In fact tiny tidal tails are present around each individual clump, but are  extremely difficult to detect.}.  
 The dynamical evolution of such  distant mergers is governed by the mutual interaction between the clumps, preventing the formation of major tidal tails.
So at high redshift, a determination of the merger rate based on the detection of tidal perturbations and  asymmetries  is extremely difficult, and not only because  low surface-brightness collisional debris would be difficult to detect in the distant Universe.

\subsection{Mass assembly of galaxies: from the present to the past: galactic archeology}
\label{sect:assembly}

In the nearby Universe, the fraction of galaxies that are involved in a major tidal interaction is believed to be very small:  a few percent. See however on Figure~\ref{fig:M31} the stellar bridge between Andromeda and M33. The debris, of extremely low surface brightness, are remnants of a tidal interaction between two of the most massive galaxies in our Local Group. Therefore the evaluation of the merger rate based on statistics on  galaxy pairs should consider systems like M31/M33 as  pairs even-though the two galaxies have a spatial separation has high  as 220~kpc.
On-going deep surveys of nearby galaxies regularly uncover very faint collisional debris. Without the observational surface brightness limit, wouldn't all  galaxies  be  connected by faint stellar bridges? A lesson to be taken away is that the fraction of tidally interacting systems might have been largely underestimated as it depends on the depth of the images used to estimate it.

\articlefigure[width=10cm]{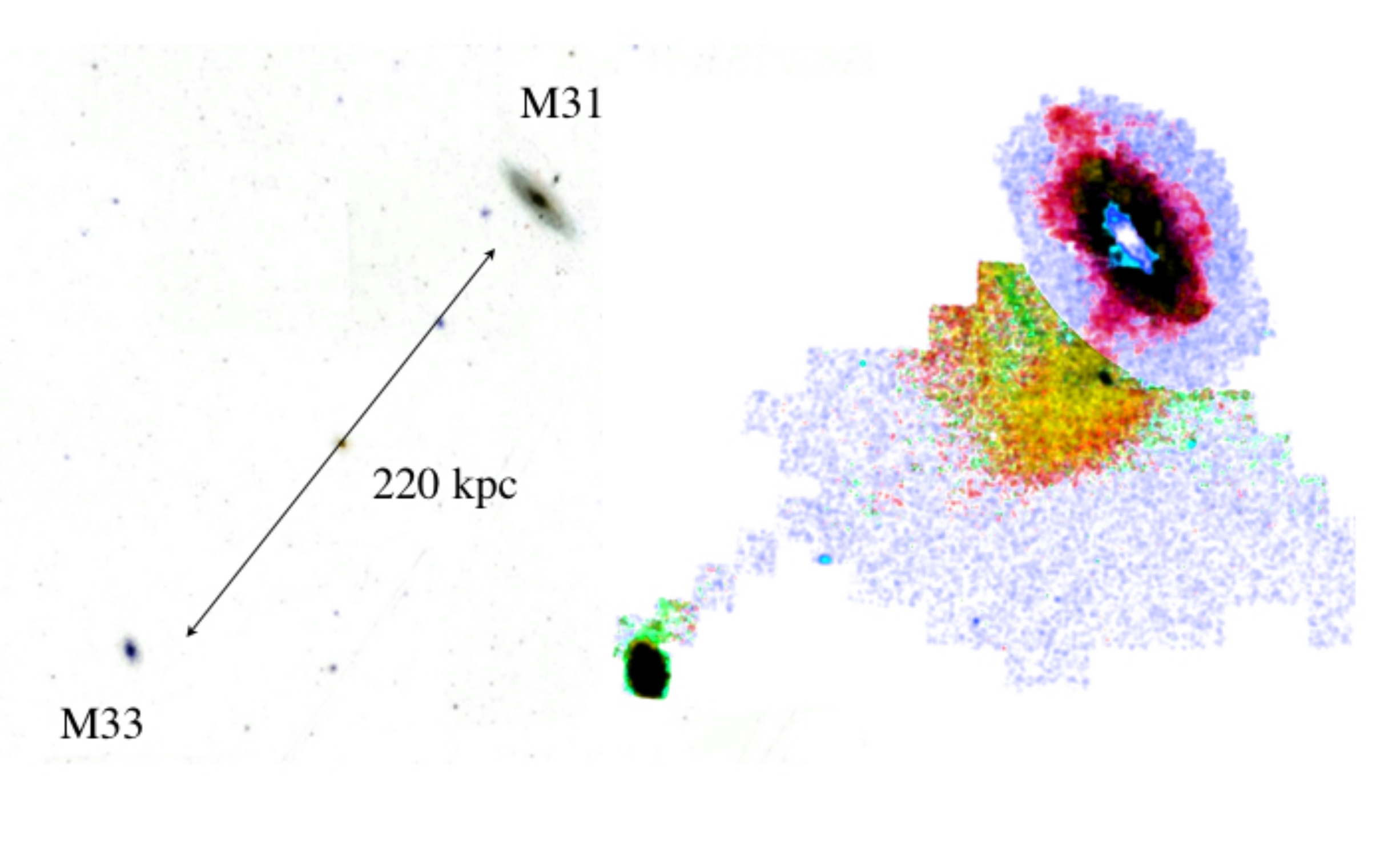}{fig:M31}{Stellar structures in the field of  M31 and M33, as seen by SDSS ({\it left}), and the CFHT/PAndAS survey ({\it right}) (courtesy R. Ibata).}  

To summarize what we have so far learned:  tidal tails and other debris keep the memory of past collisions for several Gyrs, although 
they  fade with time at a rate that may be estimated with numerical models.  Their shape and content constrain the type of collisions: minor vs major, dry vs wet.
Thus, they may in principle be used to reconstruct the past mass assembly of galaxies. 
Conversely, the absence of any collisional debris gives precious information. Mergers are not the only way galaxies grow and form bulges. Accretion by cold flows that fuel  the   distant clumpy galaxies is an alternative process that does not leave any trash in the environment. 

Therefore  with the availability of extremely deep images, galactic archeology that has started in the Local Group galaxies with stellar counts,   may be pursued on more distant systems using the now visible diffuse emission of tidal tails.

\section{Conclusion}

Should we conclude that collisional debris tell us about almost Everything You  Wanted to Know about the Universe? Alas, I Would be Afraid to claim so. 
Studies of tidal tails have not yet reached an area of precision cosmology: due to their faintness, the properties of debris remain difficult to measure. Even having  reliable photometric measurements  remain  a challenge. The interpretation of the shape of tidal tails appears rather subjective (like this contribution). For instance, how sure are we that a specific tail is really a due to a major merger and is not a stellar stream from a disrupted satellite, despite what was presumptuously said in Sect.~\ref{sect:maj-min}? Determining its color and metallically  should give further hints, but is  still very difficult.   
Furthermore, the constrains we get from the collisional debris on the structure of galaxies, distribution of dark matter, laws of physics, rely for most of them on the interpretation of numerical simulations that have their own issues: assumptions on initial conditions, limited resolution, sub-grid physics, etc...
To be more optimistic, some of these challenges will soon be addressed: the dramatic improvement in the numerical recipes, and the availability of   extremely deep images for a large number of galaxies, providing meaningful statistical data on the frequency of collisional debris will be key improvements in the near future.
   Meanwhile,  I  hope that this contribution will at least have helped to convince the Reader that we cannot say anything about tidal tails....


\bibliography{duc}

\end{document}